\documentclass[aps,prb,superscriptaddress,twocolumn,floatfix,10pt,longbibliography]{revtex4-2}

\usepackage[english]{babel}
\usepackage{lipsum}

\usepackage{multirow}
\usepackage{diagbox}
\usepackage{wasysym}
\usepackage{overpic}

\usepackage{amsmath}
\usepackage{graphicx}
\usepackage{tikz}
\usepackage{natbib}
\usepackage[colorlinks=true, allcolors=blue]{hyperref}
\usepackage{appendix}

\usepackage{tabularx}

\begin{document}

\title{Fragile topology for six-fold rotation symmetry indicated by the concentric Wilson loop spectrum}

\author{Xinyang Li}
\affiliation{Institute of Theoretical Physics, Utrecht University, Utrecht, 3584 CC, Netherlands}
\author{Lumen Eek}
\affiliation{Institute of Theoretical Physics, Utrecht University, Utrecht, 3584 CC, Netherlands}
\author{Jasper van Wezel}
\affiliation{Institute for Theoretical Physics Amsterdam, University of Amsterdam, Science Park 904, 1098 XH Amsterdam, The Netherlands}
\author{Cristiane Morais Smith}
\affiliation{Institute of Theoretical Physics, Utrecht University, Utrecht, 3584 CC, Netherlands}

\date{\today}

\begin{abstract}
We investigate topological phase transitions for the Haldane and Kane-Mele model in a lattice with $p6$ symmetry, which consists of triangles and hexagons arranged in a two-dimensional geometry. For the Haldane model, which breaks time-reversal symmetry, we calculate the Chern number using a multi-band non-Abelian Wilson loop formalism. By varying the hopping parameters in the triangles and hexagons independently, a large variety of topological phases emerge. In the presence of a next-next-nearest neighbor hopping, the phase diagram becomes even richer, with regions exhibiting high Chern numbers. Then, we consider the Kane-Mele model, for which time-reversal symmetry is preserved, and calculate the number of $\pi$-crossings in the Concentric Wilson Loop Spectrum (CWLS). This method is appropriate to determine the topological invariant for systems hosting time-reversal and rotational symmetry, but lacking all other symmetries. According to a classification based on $K$-theory, the CWLS invariant reveals topological properties even when more conventional invariants fail to detect them. The formalism was previously successfully applied to systems with 3- and 4-fold symmetry. Here, we surprisingly find that for the 6-fold-symmetry model investigated, the topology identified by this invariant is fragile, therefore questioning the claim that this should be the strong invariant missing in a complete classification of topological insulators. 

\end{abstract}

\maketitle
\section{Introduction}
The discovery of topological insulators has revealed an unexpected state of matter, which is insulating in the bulk but conducting at the edges~\cite{vonKlitzing:1980pdk,PhysRevLett.61.2015,Kane_2005,Bernevig_2006,doi:10.1126/science.1148047,Hasan_2010}. The edge conductivity is given by an integer number of the quantum of conductance, and this integer is a topological invariant~\cite{Thouless:1982zz}, connected to the topology of the bulk bands via the bulk-boundary correspondence~\cite{PhysRevLett.71.3697}.
In the non-interacting case, the possible types of topological phases were classified in the so-called ten-fold way, based on the spectral symmetries of time-reversal, charge conjugation, and sublattice (or chiral) symmetry~\cite{PhysRevB.55.1142,Kitaev_2009}. Later, this classification has been extended to include lattice symmetries~\cite{PhysRevLett.106.106802,Slager_2012,Kruthoff_2017,Po2017,Kruthoff_2019,Bradlyn2019}, non-Hermitian models~\cite{PhysRevX.9.041015,PhysRevLett.121.086803,Coulais2020}, and Floquet systems~\cite{Rechtsman_2013,PhysRevB.82.235114}.
Within this general setting, it has been proposed, based on K-theory, that the so-called concentric Wilson loop spectrum (CWLS) contains a topological invariant that is unique to time-reversal symmetric systems that are invariant under rotations but break all other lattice symmetries~\cite{Henke_2021}. This invariant was shown to yield corner states in a particular 3-fold symmetric model, and to mark topological phase transitions in a particular 2-fold symmetric model~\cite{Henke_2021}. However, these studies remained restricted to particular cases with triangular and square geometries.

Here, we introduce a model with $6$-fold rotation symmetry and map out its topological phase transitions both in the presence and absence of time-reversal symmetry. A rich phase diagram is found, containing phases with non-trivial values of the Chern number when the system lacks time-reversal invariance~\cite{Thouless:1982zz,PhysRevLett.61.2015}. In the presence of time-reversal invariance, on the other hand, non-trivial values for the FKM or $Z_2$ invariant~\cite{Kane_2005,PhysRevLett.106.106802}, as well as the concentric Wilson loop invariant are identified. Contrary to expectation~\cite{Henke_2021}, we find that the concentric Wilson loop invariant is an indicator of fragile topology~\cite{Po2018fragiletopology,Bradlyn2019}. Indeed, it characterizes the topology of isolated pairs of bands related by time-reversal symmetry, but its value can change under hybridization with trivial occupied bands. 

The outline of this paper is the following. In Sec. II, we present the Haldane model, which breaks time-reversal symmetry, and its connection to the $p6$ tight-binding model that we focus on. In addition to the nearest-neighbor (NN) hopping, we consider a complex next-nearest neighbor (NNN) hopping. A variety of band gaps and Dirac-like band crossings is shown to emerge upon varying the system parameters. In Sec. III, we characterize the topological phases by calculating the Chern number for several phases, and we study the effect of a real next-next-nearest neighbor (NNNN) hopping parameter in Sec. IV. Finally, the quantum spin Hall effect, which occurs in the presence of time-reversal symmetry, is presented alongside the concentric Wilson loop analysis in Sec. V, and our conclusions are drawn in Sec. VI. 

\section{The $p6$ tight-binding model}
\begin{figure}[tbh]
\centering
\includegraphics[width=0.8\linewidth]{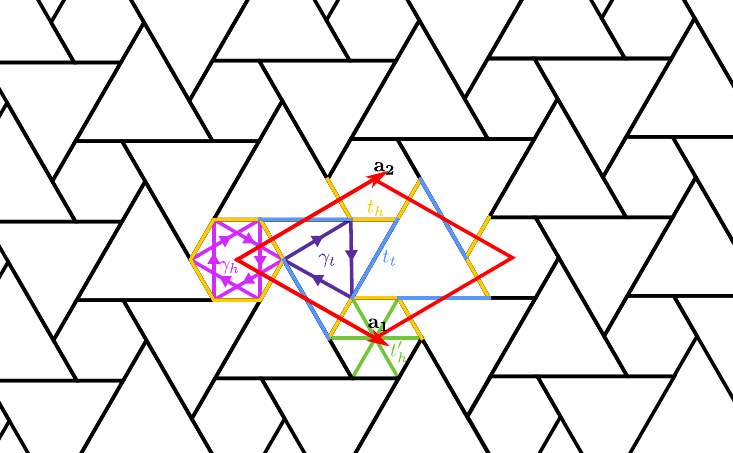}
\caption{The $p6$ lattice with five types of hoppings: $t_{h}$ (yellow lines), $t_{t}$ (blue lines), $\gamma_{h}$ (pink lines), $\gamma_{t}$ (purple lines), and $t_{h}'$ (green lines).}
\label{fig:lattice}
\end{figure}

In this work, we investigate generalizations of the Haldane~\cite{PhysRevLett.61.2015} and Kane-Mele models~\cite{Kane_2005} on the two-dimensional lattice shown in Fig.~\ref{fig:lattice}. The lattice consists of hexagons and triangles, and is designed to have six-fold ($p6$) rotation symmetry, and no other spatial symmetries. For now, we consider a spinless system with inequivalent Haldane-type hoppings in the hexagons and triangles. The Hamiltonian $H$ corresponding to the system reads
\begin{align}\label{eq:H}
    H &=  t_{h}\sum_{\langle ij\rangle \in\hexagon}c_i^\dagger c_j  + t_{t}\sum_{\langle ij\rangle \in \triangle}c_i^\dagger c_j + t^{\prime}_{h}\sum_{\langle\langle\langle ij\rangle\rangle\rangle \in \hexagon}c_i^\dagger c_j \nonumber \\ &+ i\gamma_{h}\sum_{\langle\langle ij\rangle\rangle \in \hexagon}v_{ij}c_i^\dagger c_j
     + i\gamma_t\sum_{\langle\langle ij\rangle\rangle \in \triangle}v_{ij}c_i^\dagger c_j,
\end{align}

where $c$ and $c^\dagger$ are the annihilation and creation operators, respectively. The hopping parameters are represented by colors in Fig.~\ref{fig:lattice}. Here, $t_{t}$ and $t_{h}$ represent the NN hopping in the triangle (light blue lines) and in the hexagon (yellow lines). Two types of NNN Haldane couplings are considered, represented by $\gamma_{t}$ (inside the triangle, purple lines), and $\gamma_h$ (inside the hexagon, pink lines), and $v_{ij} = 1 (-1)$ for (counter)clockwise hopping. The NNNN hopping is denoted by $t^{\prime}_h$ (green lines), and the unit cell is represented by the red parallelogram in Fig.~\ref{fig:lattice}.

\begin{figure*}[htb]
\centering
\includegraphics[width=\linewidth]{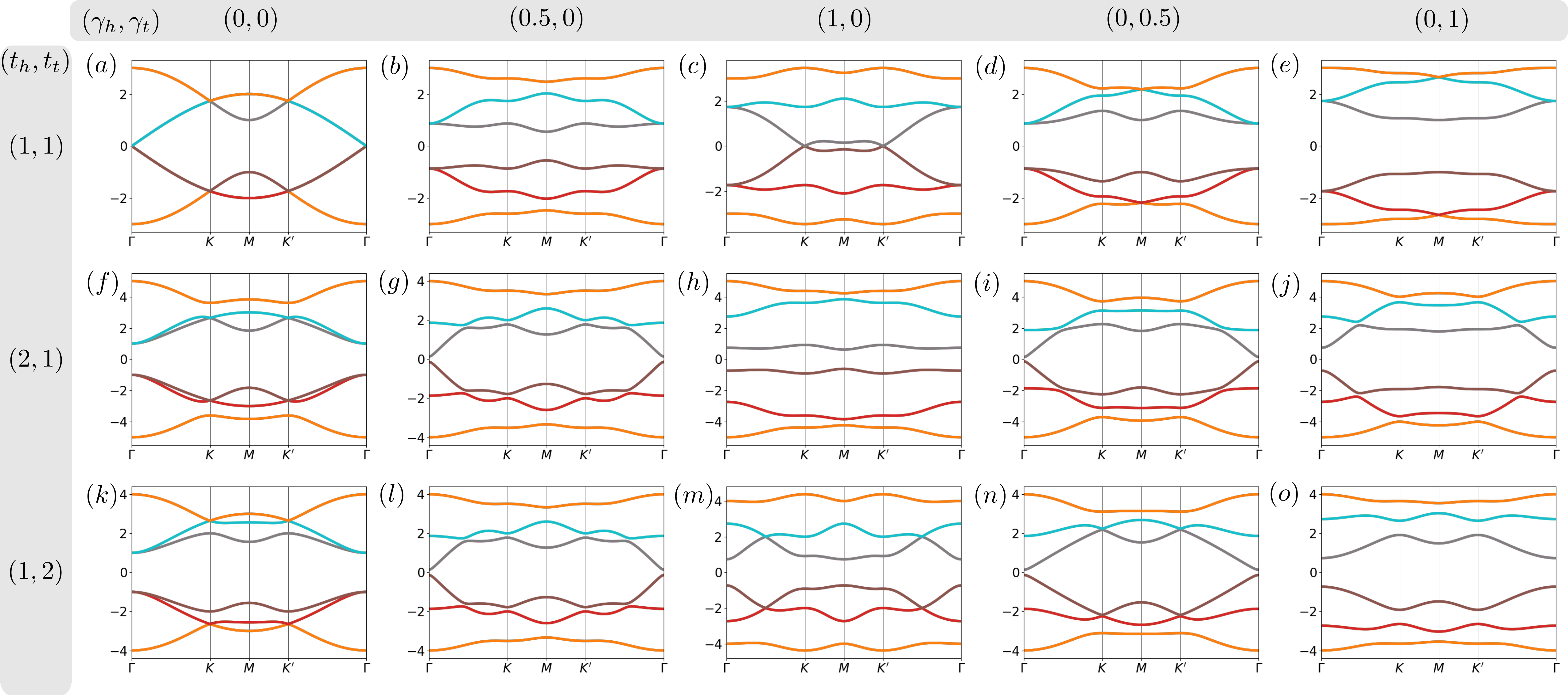}
\caption{Band structures of the $p6$ model with $\gamma_h$ and $\gamma_t$ varying from $0$ to $1$ for different values of the hopping parameters $t_h$ and $t_t$.}
\label{gammath}
\end{figure*}

The $p6$ lattice has six sites in its unit cell. Consequently, the band structure consists of six bands, labeled $E_i$, where $i$ ranges from 1 to 6. $E_1$ represents the lowest and $E_6$ the highest band. Due to chiral symmetry, the band structure is symmetric around $E = 0$. The first column in Fig.~\ref{gammath} shows the band structure of the $p6$ lattice, if only the NN hopping is included. When $t_{h}$ and $t_{t}$ are equal, the model is gapless, as shown in Fig.~\ref{gammath}(a). The second and third bands, as well as the fourth and fifth bands, are degenerate. Notably, there is a Dirac cone at zero energy at the $\Gamma$ point, and a pair of Dirac cones at finite energy at both the $K$ and $K'$ points. When $t_{h}$ and $t_{t}$ are different, a band gap opens at the $\Gamma$ point, and the degeneracies of the bands $E_2$, $E_3$, $E_4$, and $E_5$ are lifted. Specifically, when $t_{h} = 2$ and $t_{t} = 1$, the first and second bands, fifth and sixth bands are totally separated, and most of the bands are flattened, see Fig.~\ref{gammath}(b). Asymmetric Dirac cones are formed at the $K$ and $K'$ points, by the second and third, fourth and fifth bands with a small band velocity at the crossing. The structure for $t_h = 1$, $t_t = 2$ is illustrated in Fig.~\ref{gammath}(c). In this case, the Dirac cones at $K$ and $K'$ points are formed by the bands $(E_1,E_2)$ and bands $(E_5, E_6)$. 

\begin{figure}[b]
\centering
\includegraphics[width=0.6\linewidth]{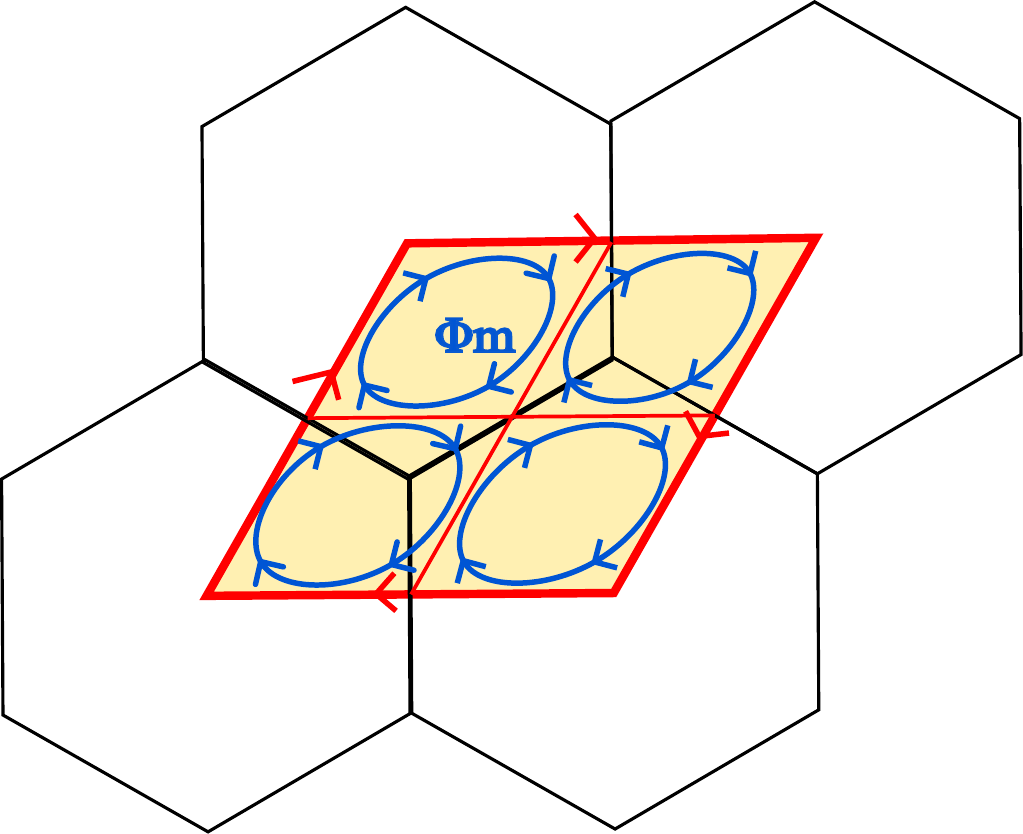}
    \caption{Schematic illustration of discrete Wilson loops. The Brillouin zone (red parallelogram) is discretized into an $N\times N$ grid (here $N=2$). For each plaquette, the Berry flux $\Phi_m$ is evaluated by computing a Wilson loop along its boundary (blue loops). The Chern number is obtained by summing the Berry flux over all plaquettes. }
    \label{ChernBZ}
\end{figure}

The second to fifth columns of Fig.~\ref{gammath} display the band structure for various Haldane couplings $\gamma_h$ and $\gamma_t$ with $t_h$ and $t_t$ held fixed. The first row in Fig.~\ref{gammath} shows the change of band structure when $t_h$, $t_f$ are fixed at 1. When Haldane terms are introduced, the degeneracy of the second and third, fourth and fifth bands is lifted. Moreover, in  Fig.~\ref{gammath}(d), when $\gamma_h = 0.5$, the Dirac cones at $K$ and $K'$ points are gapped and the degenerate Dirac cones at the $\Gamma$ point have split into higher-order crossing points. However, a larger $\gamma_h = 1$ is enough to generate new Dirac cones at the $K$ and $K'$ points, as shown in Fig.~\ref{gammath}(g). In Fig. \ref{gammath}(j), when $(\gamma_h,\gamma_t) = (0,0.5)$, a new kind of Dirac cone emerges at the $M$ point, formed by the first and second, and by the fifth and sixth bands. This kind of Dirac cone is also observed when $(\gamma_h,\gamma_t) = (0,1)$, as illustrated in Fig. \ref{gammath}(m).

In the second row, $(t_h, t_t)$ are fixed at $(2,1)$. When $\gamma_h=0.5$, the first and second bands, as well as the fifth and sixth bands form Dirac cones at the $K$ and $K'$ points, but the six bands do not intersect anywhere else in the Brillouin zone, see Fig. \ref{gammath}(e). As $\gamma_h$ reaches $1$, all bands become gaped throughout the Brillouin zone, and the third and fourth bands are flattened [Fig.~\ref{gammath}(h)]. The model is also fully gapped when $\gamma_t$ equals $0.5$ and $1$. The second and third, and the fourth and fifth bands bend toward each other between $\Gamma$ and $K$, but they never actually cross, see Fig.~\ref{gammath}(k) and Fig.~\ref{gammath}(n). 

The band structures for $(t_h, t_f) = (1,2)$ are plotted in the third row. Once $\gamma_h$ increases, the second and third, fourth and fifth bands approach each other at points between $\Gamma$ and $K$, $\Gamma$ and $K'$. Eventually, they touch each other and form Dirac cones at $\gamma_h = 1$, as shown in Fig. \ref{gammath}(i). When $\gamma_t = 0.5$, Dirac cones appear between the second and third and between the fourth and fifth bands at $K$ and $K'$, as shown in Fig.~\ref{gammath}(l). As $\gamma_t$ increases to 1, all band crossings and Dirac cones are destroyed [Fig.~\ref{gammath}(o)].

\section{Chern insulator}
In general, the effect of the topological Haldane terms is to lift degeneracies and open a gap at the $\Gamma$ point at $E = 0$. 
Furthermore, they break time-reversal symmetry, such that the induced gap may be topologically non-trivial, with its properties characterized by a Chern number. In this work, the Chern numbers are computed using a discrete Wilson loop method, where the Brillouin zone is partitioned into small plaquettes, and the total Chern number is obtained by summing the Berry flux over all plaquettes, as shown schematically in Fig.~\ref{ChernBZ}. In every plaquette, the Berry flux $\Phi_m$ through it is evaluated by a Wilson loop that circles around it,
\begin{equation} 
\Phi_m = \int_{\Gamma_m} d^2 k \, \hat{\mathbf{z}} \cdot \mathbf{b}(\mathbf{k}) = \oint_{\partial \Gamma_m} d\mathbf{k} \cdot \mathbf{a}(\mathbf{k}) , 
\end{equation}
where $\mathbf{a}(\mathbf{k})$ and $\mathbf{b}(\mathbf{k})$ are the Berry connection and curvature, respectively, and $\hat{\mathbf{z}}$ is the normal vector in the $k_z$ direction. For a sufficiently small plaquette and a single isolated band, the Berry flux can be written in terms of Bloch wave functions $u(\mathbf{k})$,
\begin{equation}
\Phi_m = i \oint_{\partial \Gamma_m} d\mathbf{k} \cdot 
\langle u(\mathbf{k}) | \nabla_{\mathbf{k}} | u(\mathbf{k}) \rangle .
\end{equation}  
For multiple bands,  it has been shown that the Berry flux $\Phi_m$ can be evaluated using a non-Abelian Wilson loop formalism instead~\cite{Fukui_2005},
\begin{equation}
    \Phi_m = \mathrm{Arg}\left\{
\prod_p \mathrm{Det}\left[
\langle u_s(\mathbf{k}_{m,p}) | u_{s'}(\mathbf{k}_{m,p+1}) \rangle
\right]
\right\}.
\end{equation}
Here, $s$ and $s'$ are the band indices, and $\mathbf{k}_{m,p}$ are the momenta along the edge of the $m$-th plaquette.

\begin{figure}
    \centering
    \includegraphics[width=\linewidth]{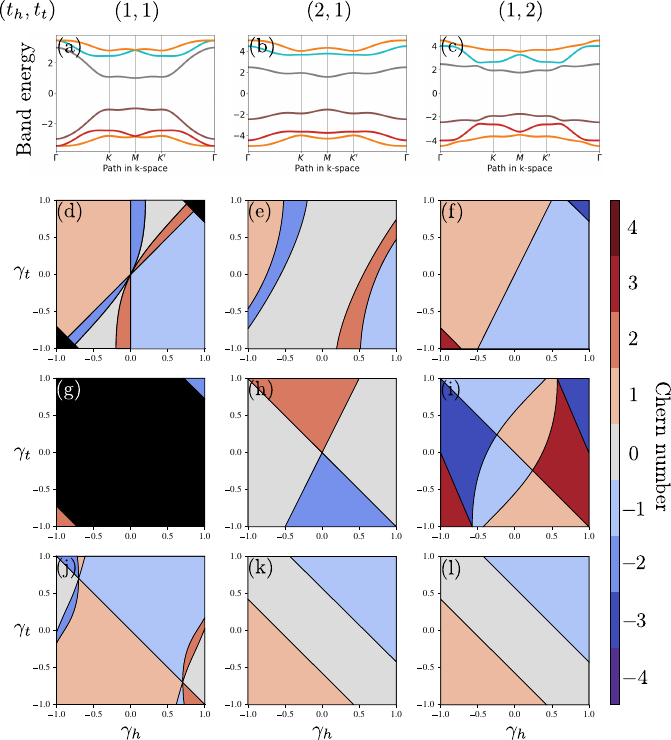}
    \caption{Topological characterization of the $p6$ lattice. (a)-(c) display the band structure for different choices of $(t_h,t_t)$ at fixed $\gamma_h=\gamma_t=1$. (d)-(f) Phase diagrams for filling $n=1$ corresponding to the above choices of $(t_h,t_t)$, for values of $\gamma_h$, $\gamma_t$ varying from $-1$ to $1$. (g)-(i) and (j)-(l) depict the same but for $2$ and $3$ filled bands, respectively. Black indicates that the system is gapless.}
    \label{fig:phasedia}
\end{figure}

\begin{figure}
    \centering
    \includegraphics[width=0.95\linewidth]{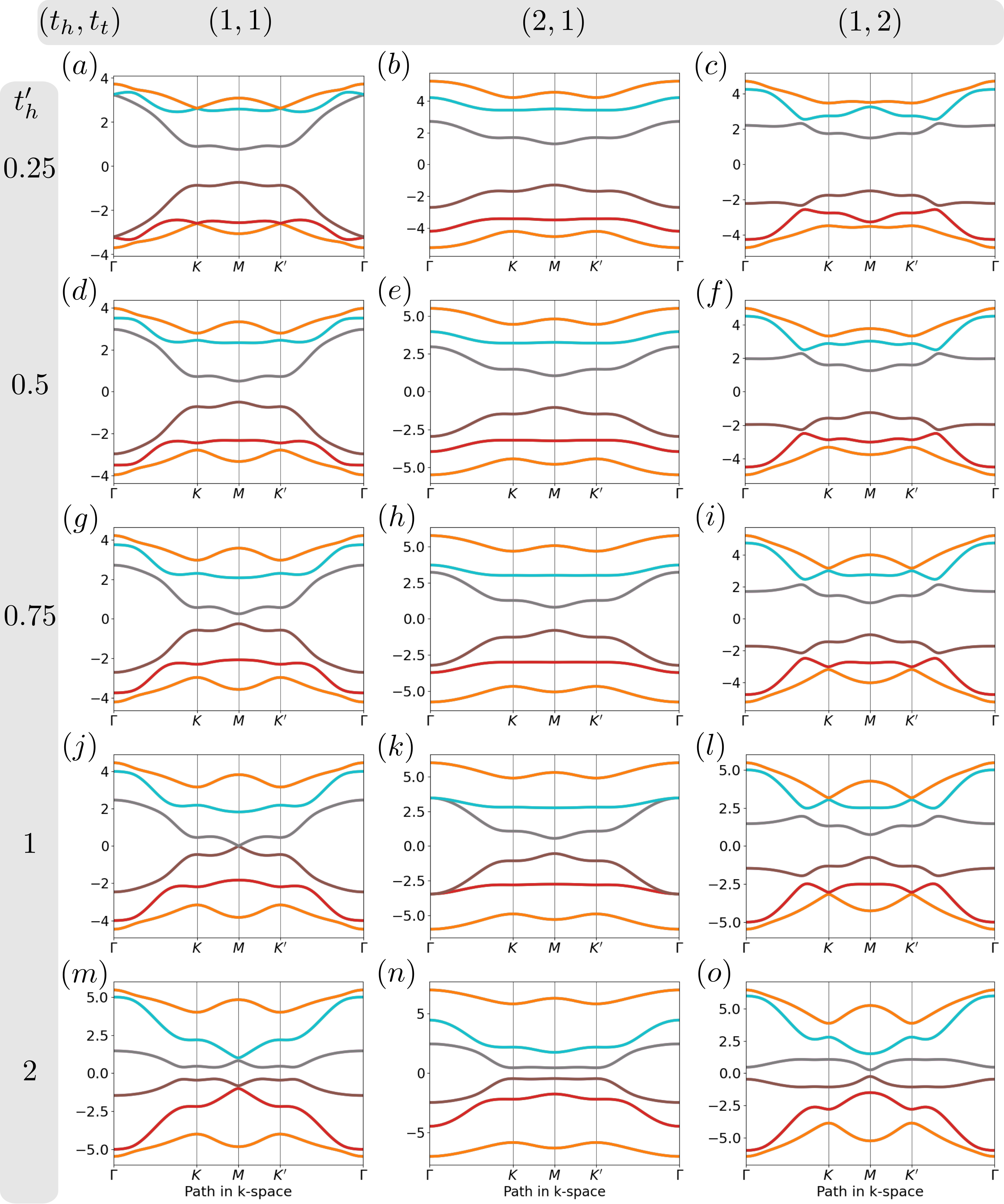}
    \caption{Band structure with NN, NNNN hopping, and $\gamma_t = \gamma_h = 1$.}
\label{th1}
\end{figure}

Finally, by summing the Berry flux over all plaquettes, we obtain the Chern number 
\begin{align} 
\nu = (2\pi)^{-1}\sum_m \Phi_m .
\end{align}
In our calculations, the Brillion zone is discretized into an $N\times N$ grid, and a grid size of $N=20$ is typically sufficient to ensure numerical convergence.

Since the band structure is symmetric with respect to $E=0$, the three upper bands do not provide independent topological information, as their Chern numbers equal those of the lower ones (up to a sign). Therefore, we only focus on the lower three bands. To this end, we calculate the Chern numbers for the first (lowest) band, the first two bands, and the first three bands of the six-band model. For each subsystem, we explicitly compute the energy gap between the highest band within the subsystem and the next band above it, ensuring that the subspace is spectrally isolated and that its Chern number is well defined. 

In each calculation, the NN hoppings $t_{h}$, $t_{t}$ are fixed, while $\gamma_{h}$ and $\gamma_t$ vary from $-1$ to $1$. For every pair $(\gamma_{h},\gamma_t)$, we compute the corresponding Chern number. The complete set of Chern numbers obtained is summarized in the first column of Table~\ref{tab:chern}.
For $t_{h} = t_{t} = 1$, the Chern numbers are restricted to values $0$, $\pm 1$, and $\pm 2$, as shown in Table~\ref{tab:chern}. Allowing for different values of $t_t$ and $t_h$ provides a route to modify the system’s topology. Figure ~\ref{fig:phasedia} illustrates the influence of different NN hopping on the topological properties of the model. The top row shows band structures for different choices of $(t_h, t_t)$ at fixed $\gamma_h=\gamma_t=1$. In Fig.~\ref{fig:phasedia}(b) and \ref{fig:phasedia}(c) new gaps can be observed upon comparison with Fig.~\ref{fig:phasedia}(a). This behavior is further captured in the phase diagrams presented in Figs.~\ref{fig:phasedia}(d)-(l) for different band subsystems. Consider, for example, 2 filled bands, i.e. the phase diagrams in Fig.~\ref{fig:phasedia}(g)-(i). When $t_h = t_t=1$, the system is mostly gapless, indicated by the large black region in Fig.~\ref{fig:phasedia}(g). Upon allowing $t_h$ and $t_t$ to differ, however, the phase diagram becomes a lot richer, with Chern numbers ranging from $-3$ to $3$, as can be observed in Figs.~\ref{fig:phasedia}(f), (i). Moreover, the different NN hopping parameters allow us to drive the system into dimerized limits ($t_t \gg t_h $ and $t_t\ll t_h$), where the system is topologically trivial. A summary of the allowed Chern numbers is presented in Tab.~\ref{tab:chern}.

\begin{table}
\centering
\caption{All possible absolute Chern numbers for nine representative parameter sets.
Each cell lists the values for the 1-, 2-, and 3-band subsystems
as $|C_1|, |C_2|, |C_3|$, for $\gamma_h,\gamma_t \in [-1,1]$.}
\label{tab:chern}
\setlength{\tabcolsep}{3pt}
\renewcommand{\arraystretch}{1.2}
\scriptsize
\begin{tabular}{c|c|c|c}
\hline\hline
\diagbox[width=5em,height=2.5em]{$\mathbf{t}$}{$t_h'$}
& \textbf{$0$} & \textbf{$0.5$} & \textbf{$1$} \\
\hline

(1,1)
& \begin{tabular}[t]{@{}l@{}}
$|C_1|:\{0,1,2\}$\\
$|C_2|:\{2\}$\\
$|C_3|:\{0,1,2\}$
\end{tabular}
& \begin{tabular}[t]{@{}l@{}}
$|C_1|:\{0,1,2\}$\\
$|C_2|:\{1,3\}$\\
$|C_3|:\{0,1,2,3,4\}$
\end{tabular}
& \begin{tabular}[t]{@{}l@{}}
$|C_1|:\{0,2\}$\\
$|C_2|:\{1,3\}$\\
$|C_3|:\{0,1,2,3,4\}$
\end{tabular}
\\
\hline

(2,1)
& \begin{tabular}[t]{@{}l@{}}
$|C_1|:\{0,1,2\}$\\
$|C_2|:\{0,2\}$\\
$|C_3|:\{0,1\}$
\end{tabular}
& \begin{tabular}[t]{@{}l@{}}
$|C_1|:\{0,2\}$\\
$|C_2|:\{0,2\}$\\
$|C_3|:\{0,1,2,3\}$
\end{tabular}
& \begin{tabular}[t]{@{}l@{}}
$|C_1|:\{0\}$\\
$|C_2|:\{\emptyset\}$\\
$|C_3|:\{0,1,2,4\}$
\end{tabular}

\\
\hline

(1,2)
& \begin{tabular}[t]{@{}l@{}}
$|C_1|:\{1,3\}$\\
$|C_2|:\{1,3\}$\\
$|C_3|:\{0,1\}$
\end{tabular}
& \begin{tabular}[t]{@{}l@{}}
$|C_1|:\{0,1,2\}$\\
$|C_2|:\{1,3\}$\\
$|C_3|:\{0,1,3\}$
\end{tabular}
& \begin{tabular}[t]{@{}l@{}}
$|C_1|:\{0,2\}$\\
$|C_2|:\{1,3\}$\\
$|C_3|:\{0,1,3\}$
\end{tabular}

\\
\hline\hline
\end{tabular}
\end{table}

\section{Effect of NNNN hopping}
Figure~\ref{th1} shows that non-zero values of the NNNN hopping $t^\prime_h$, illustrated in Fig.~\ref{fig:lattice} by green lines, can also drive topological transitions. Indeed, it was shown earlier for simpler lattices that longer-range hoppings can drive topological phase transitions \cite{beugeling_topological_2012}. Here, this is also observed. For example, in the second column of Fig.~\ref{th1}, all bands are separated except in Fig.~\ref{th1}(k), marking a phase transition. For $t_t = t_h =1$, the bands also get more isolated as $t^{\prime}_h$ increases, but bands $(E_3, E_4)$ eventually form a Dirac cone at the M point, at $t^{\prime}_h = 1$ thus signaling a phase transition [see Fig.~\ref{th1}(j)]. This Dirac cone disappears when $t^{\prime}_h$ reaches $2$, and new Dirac cones are formed by band pairs $(E_2, E_3)$ and $(E_4, E_5)$, as shown in Fig.~\ref{th1}(m). In the third column, Dirac cones are observed at $K$ and $K’$ points when $t^{\prime}_h = 0.75$ and $t^{\prime}_h = 1$, formed by bands $(E_1, E_2)$ and $(E_5, E_6)$, but they gap out for $t_h'=2$. Therefore, the variety of tunable parameters in the $p6$ lattice leads to a very rich behavior in terms of topological phases and phase transitions.

\begin{figure}
    \centering
    \includegraphics[width=\linewidth]{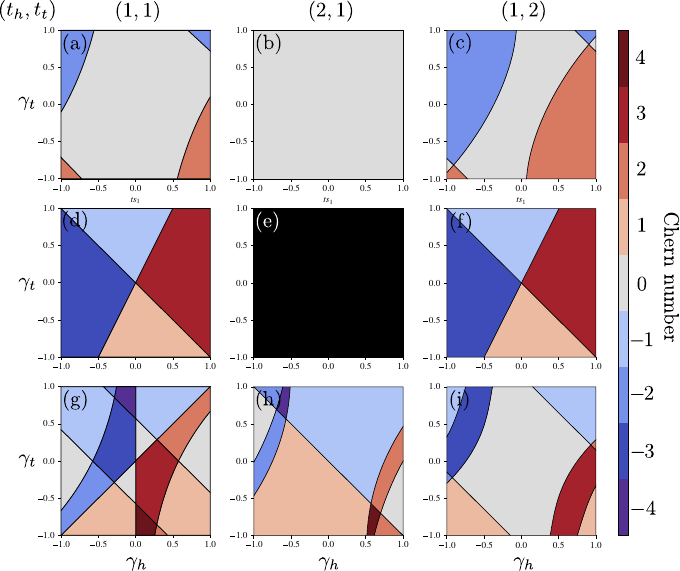}
    \caption{Topological characterization of the $p6$ lattice in the presence of NNNN hopping $t_h' = 1$. (a)-(c) Phase diagrams for filling $n=1$, corresponding to the indicated choices of $(t_h,t_t)$. (d)-(f) and (g)-(i) depict the same but for $2$ and $3$ filled bands, respectively. Black indicates that the system is gapless. In the presence of a finite $t'_h$, phases with Chern number up to $\pm4$ appear in the phase diagram.}
    \label{fig:phasediaNNNN}
\end{figure}

Moreover, considering nonzero values of $t_{h}'$ allows for higher Chern number values, such as $\pm 3$ and $\pm 4$. Indeed, already a small value of $t_h'$ is enough to open band gaps at high-symmetry points, giving rise to more topological phases (see e.g. the top row of Fig.~\ref{th1}). Figure~\ref{fig:phasediaNNNN} shows phase diagrams for the same parameters as chosen in Fig.~\ref{fig:phasedia}(d)-(l) but for $t'_h=1$ instead of zero. These phase diagrams illustrate the Chern number as $\gamma_{h}$ and $\gamma_t$ vary from $-1$ to $1$, for several values of $t_{h}$ and $t_{t}$. In the absence of NNNN hopping ($t^{\prime}_h=0$), the Chern number only takes values to be $0$, $\pm 1$, $\pm 2$ for equal NN hopping and values up to $\pm 3$ if the NN hoppings for the triangle and the hexagon differ in value. The introduction of NNNN hopping further enriched the topological phase diagram with the appearance of phases with Chern number $\pm 4$, as shown in Fig.~\ref{fig:phasediaNNNN}(g),(h).
\section{Quantum Spin Hall effect}
By taking the Hamiltonian defined in Eq.\eqref{eq:H} and combining it with a time-reversed copy, one obtains a (spinfull) Kane-Mele type model described by the Hamiltonian
\begin{align}\label{eq:HKM}
    H &=  t_{h}\sum_{\langle ij\rangle \in\hexagon}c_i^\dagger c_j  + t_{t}\sum_{\langle ij\rangle \in \triangle}c_i^\dagger c_j + t^{\prime}_{h}\sum_{\langle\langle\langle ij\rangle\rangle\rangle \in \hexagon}c_i^\dagger c_j \nonumber\\ &+ i\gamma_{h}\sum_{\langle\langle ij\rangle\rangle \in \hexagon}v_{ij}c_i^\dagger s_z c_j
     + i\gamma_t\sum_{\langle\langle ij\rangle\rangle \in \triangle}v_{ij}c_i^\dagger s_z c_j.
\end{align}
Here, all parameters are the same as in Eq.~\eqref{eq:H}, except that $\gamma_t$ and $\gamma_h$ now correspond to intrinsic spin-orbit coupling (SOC) strengths. Furthermore, $s_z$ is the $z$-Pauli matrix.

According to the classification based on K-theory~\cite{Kruthoff_2019}, the topological properties of this model are described by three independent topological invariants: the $Z_2$ invariant~\cite{Kane_2005}, the Lau-Brink-Ortix (LBO) invariant~\cite{PhysRevB.94.165164}, and the more recently proposed $\pi$-crossings in the Concentric Wilson Loop Spectrum ~\cite{Henke_2021}. The CWLS is defined for topological insulators with time-reversal symmetry and rotational symmetry~\cite{Henke_2021}. It has been previously applied to systems with three and four-fold rotational symmetry~\cite{Henke_2021}. Here, we establish its relevance to situations with six-fold rotational symmetry by applying it to the $p6$ model.

\begin{figure*}
    \centering
    \includegraphics[width=\linewidth]{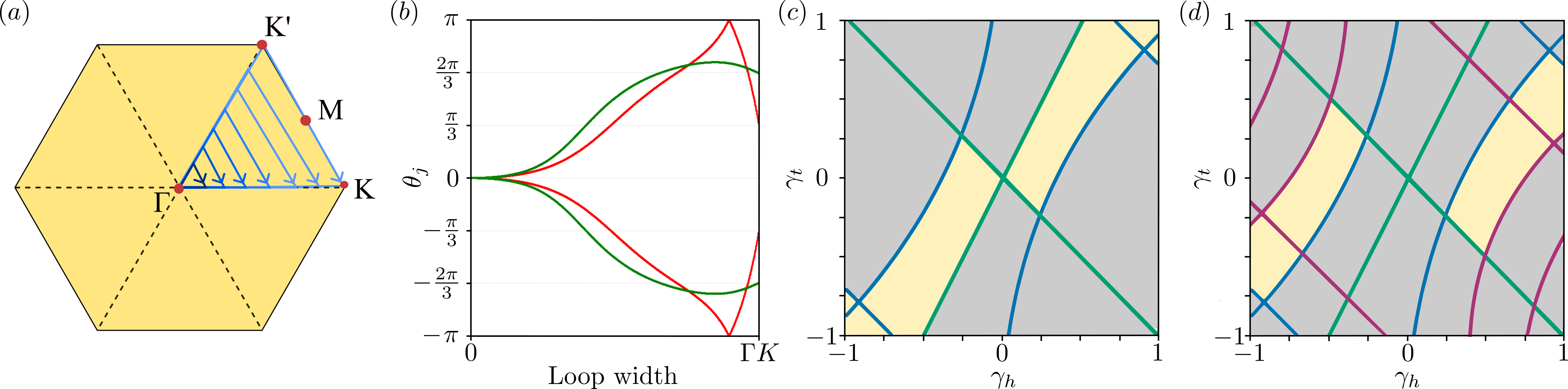}
    \caption{(a) Concentric Wilson loops in a BZ with six-fold rotational symmetry, where the  Wilson loops only enclose $1/6$ of the BZ. (b) CWLS spectra of the second-lowest Kramers pair at two representative points: $(\gamma_h, \gamma_t) = (-1, -0.8)$ (green lines) and $(\gamma_h, \gamma_t) = (-0.85, -1)$ (red lines) for $(t_h, t_t,t_h')=(1,1,0)$. Both points lie in a parameter region with the same gap structure: the internal gap between $(E_2, E_3)$ closes, but the gap between $(E_4, E_5)$ stays open. (c,d) Phase diagrams obtained by summing the number of $\pi$-crossings of the lowest two (c) and three (d) Kramers pairs. In both images, $(t_h, t_t, t'_h) = (1, 2, 1)$. Gap-closing lines between $(E_2, E_3)$, $(E_4, E_5)$, and $(E_6, E_7)$ are indicated by blue, green, and red lines, respectively. The yellow regions correspond to a summed $w \bmod 2 = 1$, while the gray regions correspond to $w \bmod 2 = 0$.}
    \label{fig:CWLScombined}
\end{figure*}

 The CWLS is defined by a family of Wilson loops starting at the BZ center and expanding to enclose fractional sectors of size $1/n$ of the BZ, consistent with the $n$-fold rotational symmetry. In the presence of time-reversal symmetry, every state $|\psi\rangle$ has a time-reversal partner $T|\psi\rangle$ with the same energy, leading to the formation of Kramers pairs. Therefore, it is natural to consider the $U(2)$ concentric Wilson loops for individual Kramers pairs. Eigenvalues of the Wilson loop are then plotted against loop radius; a so-called $\pi$-crossing occurs when the phases of the Wilson loop eigenvalues cross $\pi$, and the parity of such crossings yields the CWLS invariant. The endpoint of the CWLS is also quantized: the point at which the Kramers pair ends multiplied by $n$ and divided by $2\pi$ equals the value of the $Z_2$ invariant for the single Kramers pair. The CWLS with $C_6$ rotational symmetry in the BZ is illustrated in Figure \ref{fig:CWLScombined}(a). 

In the presence of time-reversal symmetry, the $p6$ model has 12 bands, and they are symmetric with respect to $E = 0$, due to the chiral symmetry. Therefore, we focus on the two, four, and six lower band subsystems. 

We systematically map the three time-reversal symmetric invariants under various parameter choices. The $Z_2$ invariant is computed using the parallel Wilson loop approach~\cite{Yu_2011}, and the LBO invariant is associated with the eigenvalues of Wilson loops along $k_y = 0$. The $Z_2$ invariant is defined for the full occupied subspace, and thus all the bands below the Fermi level are taken into account in the calculation. In contrast, the parity of $\pi$-crossing of the CWLS is defined for every isolated Kramers pair, and it relies on the presence of finite band gaps below and above it. 
This is shown explicitly in Fig. \ref{fig:CWLScombined}(b), which displays the CWLS of two sets of parameters that are connected by a closing of the internal gap between $(E_2,E_3)$, while the gap between $(E_4,E_5)$ stays open.

If we sum the number of $\pi$-crossings of all CWLS for occupied Kramers pairs together, the resulting quantity reflects the influence of all internal gaps within the subspace. As an illustration, Figs.~\ref{fig:CWLScombined}(c) and (d) shows results summing the number of $\pi$-crossings of the lowest two Kramers pairs [Fig.~\ref{fig:CWLScombined}(c)], as well as those of the lowest three Kramers pairs [Fig.~\ref{fig:CWLScombined}(d)]. In both cases, the sum is taken modulo 2. 

Surprisingly, the resulting phase diagrams are partitioned by all gap-closing lines inside the band subspace, rather than only the one above the subspace summed over. This demonstrates that the sum of CWLS $\pi$-crossings is in fact a fragile invariant~\cite{Po2018fragiletopology,Bradlyn2019}. This means that, its value is well defined, as long as the occupied Kramers pairs stay spectrally separated, but it may change under hybridization of occupied Kramers pairs.

The full set of possible phases for all parameter sets considered are summarized in Tab.~\ref{TRSphase}. In this table, the LBO invariant remains trivial. Within every box, every line represents a Kramers pair, we only consider the first 3 Kramers pairs as the other 3 behave symmetrically due to TRS. Since the CWLS is $U(2)$ it is only calculated if the given Kramers pair is separated from the others, if this is not possible, we write $\emptyset$. $w=1$ $(w=0)$ indicates that the CWLS winds (does not wind) and $w=1^*$ indicates that the winding only occurs at the endpoint of the loop. The numbers in the brackets represent the possible endpoints of the CWLS with $\{n\}$ indicating $\pm n\pi/3$ as the endpoint.

\begin{figure}
\centering
\includegraphics[width=\linewidth]{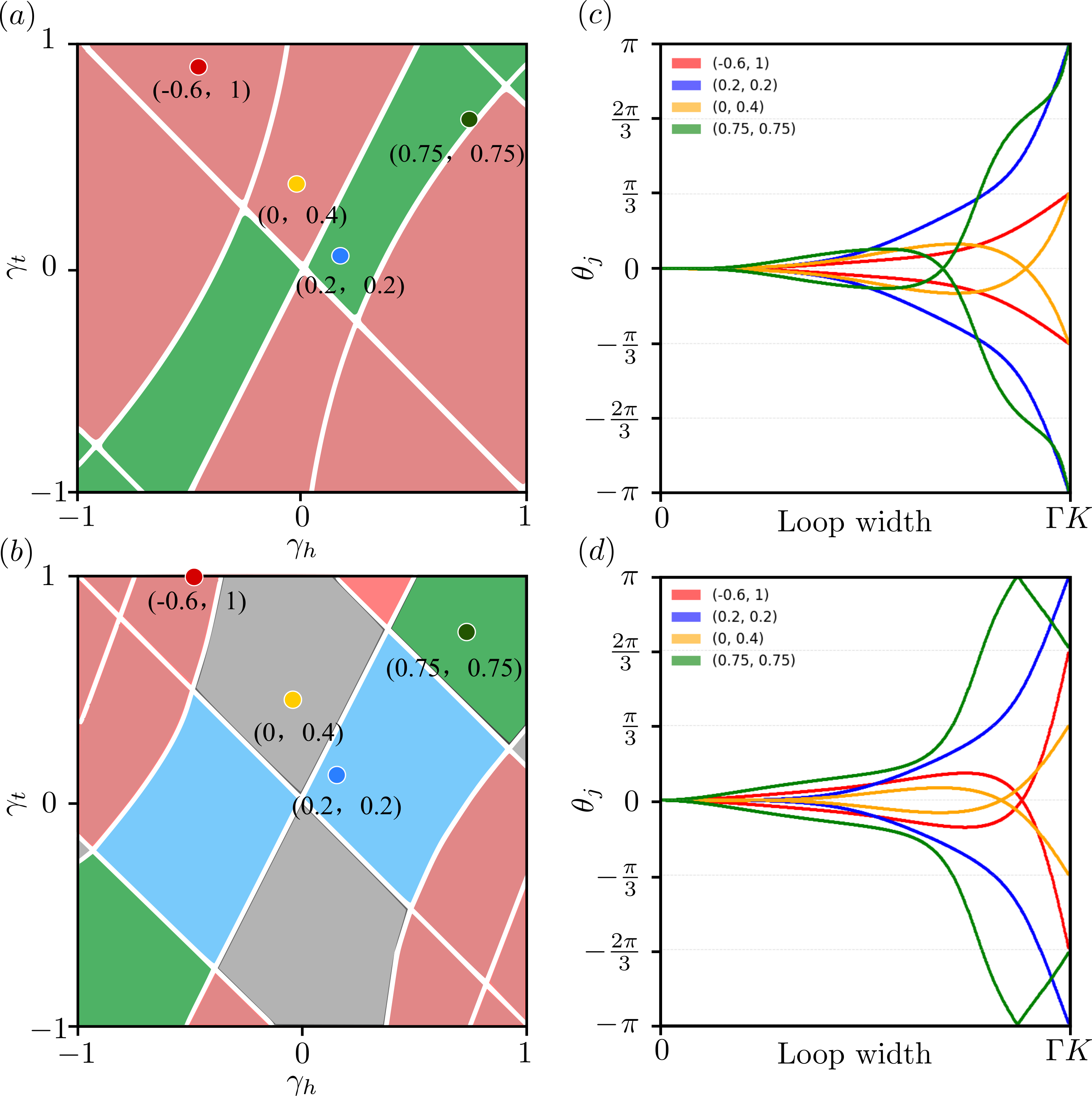}
\caption{(a),(c) Phase diagrams of the four-band and six-band subsystems parameterized by $\gamma_h$ and $\gamma_t$, at fixed $(t_h, t_t, t'_h) = (1, 2, 1)$.
The phases are characterized by the $Z_2$ invariant and the CWLS (denoted below as $w$).
Different colors indicate distinct topological phases: pink regions correspond to $(Z_2, w) = (1,0)$, green to $(1,1)$, blue to $(0,1)$, and gray to the trivial phase.
White lines indicate gap-closing lines. In (a), band closings occur between $(E_2, E_3)$ and $(E_4, E_5)$; in (b), between $(E_4, E_5)$ and $(E_6, E_7)$.
Four representative points are marked by colored dots. (c), (d) CWLS calculated at the representative points indicated in (a) and (b).}
\label{fig:121phase}
\end{figure}

To exemplify the trivial and non-trivial CWLS occurring in this model, we present phase diagrams for subsystems consisting of the lowest four and six bands in Fig.~\ref{fig:121phase}. In this example, we scan $\gamma_h$ and $\gamma_t$ at fixed ($t_h$, $t_t$, $t’_h$) = $(1, 2, 1)$, and systematically study the topological phases with four- and six-band subsystems. The topological phases are characterized by the $Z_2$ invariant and CWLS (labeled as $w$ in the figure). Each of them can take the value of 1 or 0, yielding four distinct topological phases. The different phases are depicted in Figs.~\ref{fig:121phase}(a) and (b), labeled with different colors. The white lines indicate gap-closing lines below and above the relevant Kramers pair. For every phase in Figs.~\ref{fig:121phase}(a) and (b), a representative point is chosen and highlighted. Their CWLS are calculated and illustrated in Figs.~\ref{fig:121phase}(c) and (d), respectively.

\begin{table}
\centering
\caption{All possible phases with TRS, where $t'_h$ takes $0$, $0.5$, $1$, and ($t_h$,$t_t$) are set to be $(1,1)$, $(1,2)$ and $(2,1)$. $w$ represents the CWLS invariant. In this table, the LBO invariant remains trivial. Invariants are only evaluated for individual Kramers pairs, $\emptyset$ indicates that the Kramers pairs can not be separated.}
\label{TRSphase}
\setlength{\tabcolsep}{1pt}
\renewcommand{\arraystretch}{1.2}
\scriptsize
\resizebox{0.975\columnwidth}{!}{%
\begin{tabular}{c|c|c|c}
\hline\hline
\diagbox[width=5em,height=2.5em]{$\mathbf{t}$}{$t_h'$}
& \textbf{$0$} & \textbf{$0.5$} & \textbf{$1$} \\
\hline

(1,1)
& \begin{tabular}[t]{@{}l@{}}
$w=0\phantom{,1^*}$; $\{0,1,2\}\phantom{,3}$\\
$w=\emptyset\phantom{,1^*}$; $\{\}$\\
$w=0\phantom{,1^*}$; $\{1\}$
\end{tabular}
& \begin{tabular}[t]{@{}l@{}}
$w=0\phantom{,1^*}$; $\{0,1,2\}$\\
$w=0,1\phantom{^*}$; $\{1,2,3\}$\\
$w=0,1^*$; $\{0,1,2,3\}$
\end{tabular}
& \begin{tabular}[t]{@{}l@{}}
$w=0\phantom{,1^*}$; $\{0,2\}$\\
$w=0, 1^*$; $\{1,3\}$\\
$w=0, 1^*$; $\{0,1,2,3\}$
\end{tabular}
\\
\hline

(2,1)
& \begin{tabular}[t]{@{}l@{}}
$w=0\phantom{,1^*}$; $\{0,1,2\}$\\
$w=0,1\phantom{^*}$; $\{0,1,2\}$\\
$w=0,1^*$; $\{0,1,2,3\}$
\end{tabular}
& \begin{tabular}[t]{@{}l@{}}
$w=0\phantom{,1^*}$; $\{0,2\}$ \\
$w=0,1\phantom{^*}$; $\{0,2\}$\\
$w=0, 1^*$; $\{1,2,3\}\phantom{,3}$
\end{tabular}
& \begin{tabular}[t]{@{}l@{}}
$w=0\phantom{,1^*}$; $\{0\}\phantom{,1,2,3}$\\
$w=\emptyset\phantom{,1^*}$; $\{\}$\\
$w=\emptyset\phantom{,1^*}$; $\{\}$
\end{tabular}
\\
\hline

(1,2)
& \begin{tabular}[t]{@{}l@{}}
$w=0,1^*$; $\{1,3\}$\\
$w=0,1\phantom{^*}$; $\{0,2\}$\\
$w=0,1\phantom{^*}$; $\{0,1,2,3\}$
\end{tabular}
& \begin{tabular}[t]{@{}l@{}}
$w=0\phantom{,1^*}$; $\{0,1,2\}$\\
$w=0,1\phantom{^*}$; $\{1,2,3\}$\\
$w=0,1\phantom{^*}$; $\{0,1,2,3 \}$ 
\end{tabular}
& \begin{tabular}[t]{@{}l@{}}
$w=0\phantom{,1^*}$; $\{0,2\}$\\
$w=0,1\phantom{^*}$; $\{1,3\}$\\
$w=0,1\phantom{^*}$; $\{0,1,2,3\}$
\end{tabular}
\\
\hline\hline
\end{tabular}
}
\end{table}

\begin{table}
\centering
\caption{$Z_2$ invariants for the same parameters as in Table~\ref{TRSphase}. Invariants are only evaluated for individual pairs, $\emptyset$ indicates that the Kramers pairs can not be separated.}
\label{TRSphase_Z2}
\setlength{\tabcolsep}{15pt}
\renewcommand{\arraystretch}{1.2}
\scriptsize
\begin{tabular}{c|c|c|c}
\hline\hline
\diagbox[width=5em,height=2.5em]{$\mathbf{t}$}{$t_h'$}
& \textbf{$0$} & \textbf{$0.5$} & \textbf{$1$} \\
\hline

(1,1)
& \begin{tabular}[t]{@{}l@{}}
$Z_2=0,1$\\
$Z_2=\emptyset$\\
$Z_2=0,1$
\end{tabular}
& \begin{tabular}[t]{@{}l@{}}
$Z_2=0,1$\\
$Z_2=1$\\
$Z_2=0,1$
\end{tabular}
& \begin{tabular}[t]{@{}l@{}}
$Z_2=0$\\
$Z_2=1$\\
$Z_2=0,1$
\end{tabular}
\\
\hline

(2,1)
& \begin{tabular}[t]{@{}l@{}}
$Z_2=0,1$\\
$Z_2=0$\\
$Z_2=0,1$
\end{tabular}
& \begin{tabular}[t]{@{}l@{}}
$Z_2=0$\\
$Z_2=0$\\
$Z_2=0,1$
\end{tabular}
& \begin{tabular}[t]{@{}l@{}}
$Z_2=0$\\
$Z_2=\emptyset$\\
$Z_2=\emptyset$
\end{tabular}
\\
\hline

(1,2)
& \begin{tabular}[t]{@{}l@{}}
$Z_2=1$\\
$Z_2=1$\\
$Z_2=0,1$
\end{tabular}
& \begin{tabular}[t]{@{}l@{}}
$Z_2=0,1$\\
$Z_2=1$\\
$Z_2=0,1$
\end{tabular}
& \begin{tabular}[t]{@{}l@{}}
$Z_2=0$\\
$Z_2=1$\\
$Z_2=0,1$
\end{tabular}
\\
\hline\hline
\end{tabular}
\end{table}

As mentioned before, the value at the rightmost end of the CWLS reflects the contributions of each Kramers pair to the total $Z_2$ invariant. Multiplying the endpoint by six and dividing by $2\pi$ yields a binary value, either 0 or 1. This value corresponds to a trivial or non-trivial contribution associated with the chosen Kramers pair. Alternatively, the contribution of a given Kramers pair to the total $Z_2$ invariant can be extracted by comparing the total $Z_2$ indices evaluated by the Berry flux method for subsystems that include and exclude this Kramers pair. For instance, the contribution of the second Kramers pair is obtained from the difference between the total $Z_2$ invariants of the six-band and four-band subsystems. The $Z_2$ values are listed and compared in Table \ref{TRSphase_Z2}, demonstrating consistency between the CWLS and Berry flux calculations.

\section{Conclusions}
In this work, we mapped out the topological phases of a tight-binding model on a lattice with 6-fold rotational symmetry. The lattice consists of hexagons and triangles arranged in two dimensions, reminiscent of the kagome lattice. The model includes both real NN and NNNN hopping, besides a complex NNN hopping to account for a flux in the plaquettes (analogous to the Haldane model) or an intrinsic SOC (as in the Kane-Mele model). We derived the phase diagram as a function of various parameters, and showed that the model is sufficiently rich to host a myriad of topologically distinct phases. 

In the Haldane model, time-reversal symmetry is broken and Chern numbers are calculated using conventional methods~\cite{Fukui_2005}. We found that large Chern numbers can arise naturally in this model, particularly in the presence of NNNN hopping in the hexagons. On the other hand, for the time-reversal symmetric Kane-Mele-like model, we calculated the concentric Wilson loop invariant, which was recently proposed to describe additional topological phases, not captured by conventional invariants~\cite{Henke_2021}. We identified phases in which the concentric Wilson loop invariant is non-trivial, thus extending the previous implementations on 3- and 4-fold symmetric models reported in the literature~\cite{Henke_2021} to a 6-fold symmetric setting. 

Unexpectedly, we find that the concentric Wilson loop invariant is an indicator of fragile topology. This is made explicit by the observation that its value can change under hybridization within the set of occupied bands. These results indicate that the concentric Wilson loop invariant of individual Kramers pairs is not the strong invariant that was suggested to exist based on K-theory arguments~\cite{Kruthoff_2019}.

Therefore, contrarily to previous believe, the identification of the missing invariant is not yet settled, and further research will be necessary for answering this puzzling question.  

\section*{Acknowledgement}
L.E and C.M.S acknowledge the research program “Materials for the Quantum Age” (QuMat) for financial support. This
program (registration number 024.005.006) is part of the
Gravitation program financed by the Dutch Ministry of
Education, Culture and Science (OCW).

\bibliographystyle{apsrev4-2}  %
\bibliography{sample}

\end{document}